\documentclass[prd,superscriptaddress,nofootinbib,amsmath,amssymb,aps,11pt]{revtex4}

\usepackage{bm}
\usepackage{amsfonts}
\usepackage{latexsym}
\usepackage[latin1]{inputenc}
\usepackage{graphicx}
\usepackage{amsmath}
\usepackage{palatino}
\usepackage{mathpazo}
\usepackage[normalem]{ulem}

\usepackage{booktabs}
\usepackage{dcolumn}

\def\jnl@style{\it}
\def\aaref@jnl#1{{\jnl@style#1}}

\def\aaref@jnl#1{{\jnl@style#1}}

\def\aj{\aaref@jnl{AJ}}                   
\def\apj{\aaref@jnl{ApJ}}                 
\def\apjl{\aaref@jnl{ApJ}}                
\def\apjs{\aaref@jnl{ApJS}}               
\def\apss{\aaref@jnl{Ap\&SS}}             
\def\aap{\aaref@jnl{A\&A}}                
\def\aapr{\aaref@jnl{A\&A~Rev.}}          
\def\aaps{\aaref@jnl{A\&AS}}              
\def\mnras{\aaref@jnl{Mon.~Not.~Roy.~Astron.~Soc.}}             
\def\prd{\aaref@jnl{Phys.~Rev.~D}}        
\def\prc{\aaref@jnl{Phys.~Rev.~C}}  
\def\prl{\aaref@jnl{Phys.~Rev.~Lett.}}    
\def\qjras{\aaref@jnl{QJRAS}}             
\def\skytel{\aaref@jnl{S\&T}}             
\def\ssr{\aaref@jnl{Space~Sci.~Rev.}}     
\def\zap{\aaref@jnl{ZAp}}                 
\def\nat{\aaref@jnl{Nature}}              
\def\aplett{\aaref@jnl{Astrophys.~Lett.}} 
\def\apspr{\aaref@jnl{Astrophys.~Space~Phys.~Res.}} 
\def\physrep{\aaref@jnl{Phys.~Rep.}}      
\def\physscr{\aaref@jnl{Phys.~Scr}}       
\def\commat{\aaref@jnl{Comm.~Math.~Phys.}}              
\def\science{\aaref@jnl{Science}}               
\def\cqg{\aaref@jnl{Classical Quant.~Grav.}}            
\def\jpcs{\aaref@jnl{JPCS}}                                     
\def\ijmpd{\aaref@jnl{Int.~J.~Mod.~Phys.~D}}                    
\def\grg{\aaref@jnl{Gen.~Relat.~Gravit.}}               
\def\rpp{\aaref@jnl{Rep.~Prog.~Phys.}}          
\def\npa{\aaref@jnl{Nucl.~Phys.~A}}        
\def\lrr{\aaref@jnl{Living Rev.~Rel.}}                   
\def\jcap{\aaref@jnl{J.~Cosmology Astropart.~Phys.}}    
\def\rmp{\aaref@jnl{Rev.~Mod.~Phys.}}   

\allowdisplaybreaks[1]

\begin{document}

	\title{Scalarized non-topological neutron stars in multi-scalar Gauss-Bonnet gravity  }

	\author{Kalin V. Staykov}
	\email{kstaykov@phys.uni-sofia.bg}
	\affiliation{Department of Theoretical Physics, Faculty of Physics, Sofia University, Sofia 1164, Bulgaria}
	
	\author{Radostina Z. Zheleva}
	\email{radostinazheleva94@gmail.com}
	\affiliation{Department of Theoretical Physics, Faculty of Physics, Sofia University, Sofia 1164, Bulgaria}


	\begin{abstract}
	In the present paper we  construct novel non-topological, spontaneously scalarized neutron stars in multi-scalar Gauss-Bonnet gravity with maximally symmetric target space, and nontrivial map $\varphi:spacetime \rightarrow target\ space$. The theory is characterized by the fact that for some classes of coupling functions the field equations allow solutions with trivial scalar field, which coincide with the general relativistic ones. For a certain range of parameters those solutions lose stability and new branches of solutions with nontrivial scalar field bifurcate from the trivial branch. For a given set of parameters, those branches are characterized by the number of zeros of the scalar field, and they are energetically more favorable than the general relativistic ones.  
	\end{abstract}

	\maketitle
	
	\section{Introduction}

The recent gravitational waves (GW) observations from different binary sources paved the pathway to the  GW and the multi-messenger astronomy. This should allow modified theories of gravity to be more extensively tested in the strong field regime in foreseeable future, and additional constraints on there parameters to be set. 

Of particular interest are the scalar-tensor theories (STT) which posse scalar degrees of freedom, and more precisely STT which are equivalent to General relativity (GR) in weak field regime but can deviate significantly in the strong field regime. Those theories are characterized by the fact that the field equations admit the GR solution in the case of trivial (vanishing) scalar field, but for some set of parameters this solution branch becomes unstable and new, energetically more favorable branches of solutions, bifurcate from it. Such theory was first proposed in \cite{Damour1993} but its parameters are now constrained by the observations so that no significant deviations from GR occur. 
Recently scalar-tensor theories with multiple scalar fields \cite{Damour1992,Horbatsch2015}, and STT with higher order curvature invariants, namely Gauss-Bonnet theory \cite{Berti2015a,Pani2011e,Yunes2011,Pani2011d}, regained attention. Both classes of theories allow for new types of compact objects \cite{Yazadjiev2019,Collodel2020,Doneva2020d,Doneva2020a} which exhibit significant deviations from GR in the strong field regime, being in the same time in correlation with the present observational constraints. In tensor-multi-scalar theory (TMST) with target space metric which allows periodic Killing flows, tensor-multi-scalar solitons \cite{Yazadjiev2019,Collodel2020} and mixed configurations of tensor-multi-scalar solitons and neutron stars (NS) \cite{Doneva2020d} were observed. In TMST with target space $\mathbb{S}^3$  the so-called topological neutron stars \cite{Doneva2018a,Doneva2020a,Danchev2020,Doneva2020c} were constructed for which the central value of the scalar field is equal to $n\pi$, where $n$ is a whole number with topological origin. In the case $n=0$ non-topological spontaneously scalarized solutions were constructed as well \cite{Doneva2020c}. The stability of some of those TMST compact objecs was studied in \cite{Doneva2020b,Falcone2021}. In addition, the scalar field in the TMST has no scalar charge, therefore to a large extend they evade the constraints from binary pulsar observations \cite{Freire2012,Antoniadis2013}. In the Gauss-Bonnet (GB) case, new type of scalarization was observed -- the so-called curvature induced scalarization   which is a result of the curvature of spacetime itself, in contrast to the standard STT case where the source of scalarization is the presence ot matter. Both neutron star and black hole scalarized solutions were constructed in this theory \cite{Doneva2018a,Doneva2018,Silva2018,Antoniou2018,Antoniou2018a,Minamitsuji2019,Silva2019,Brihaye2019,Myung2019,Hod2019}.  

In the present paper we study scalar-tensor theory with multiple scalar fields coupled to a higher order curvature invariant, namely multi-scalar Gauss-Bonnet (MSGB) gravity. The theory we study is characterized by three scalar fields which take values on a three-dimensional maximally symmetric target space, namely    $\mathbb{S}^3$,  $\mathbb{H}^3$ or  $\mathbb{R}^3$, and nontrivial map  $\varphi:spacetime \rightarrow target\ space$. Black holes with linear and exponential coupling as well as scalarized black holes with curvature induced scalarization in MSGB gravity were studied in \cite{Doneva2020}.  The MSGB theory allows high degree of freedom when constructing the models -- in the choice of the coupling function and the coupling constant, and in the choice of the target space, its metric and the map $\varphi:spacetime \rightarrow target\ space$. In addition, the scalar field has no scalar charge, like in the tensor-multi-scalar gravity, therefore there are no strong observational constraints on the free parameters in the theory.   

Similarly to the tensor-multi-scalar theories, the condition for regularity in the center of the star allows central values for the  scalar field equal to $n\pi$ for target space $\mathbb{S}^3$, and zero scalar field in the center for all three maximally symmetric spaces quoted above. In the present paper we study models with zero scalar field in the center (non-topological neutron stars in accordance with the terminology in the TMST).

The paper is constructed as follow: in Section II we present the reduced field equations for MSGB theory with matter. In Section III we present the conditions which fix the specific theory we study and the obtained numerical results. The paper ends with a Conclusion.

	\section{Multi-scalar Gauss-Bonnet gravity }

    In this paper we study neutron stars in multi-scalar Gauss-Bonnet gravity with $N$ scalar fields $\varphi=(\varphi^1,...,\varphi^N)$ which take values on a patch of a $N$-dimensional Reimannian manifold, called $target\ space$, equipped with positively defined metric $\gamma_{ab}(\varphi)$. For more mathematical details we refer the reader to \cite{Doneva2020}.
	The general form of the theory action is given by 
	
	\begin{eqnarray}
	S=&\frac{1}{16\pi G}\int d^4x \sqrt{-g} 
	\Big[R -  2g^{\mu\nu}\gamma_{ab}(\varphi)\nabla_{\mu}\varphi^{a}\nabla_{\nu}\varphi^{b} - V(\varphi) 
	+ \lambda^2 f(\varphi){\cal R}^2_{GB} \Big] + S_{{\rm{matter}}}\left(g_{\mu\nu},\varphi\right),\label{eq:quadratic}
	\end{eqnarray}
	where $R$ is the Ricci scalar with respect to the spacetime metric $g_{\mu\nu}$,   $V(\varphi)$ is the potential of the scalar fields. The coupling function  $f(\varphi)$ depends only on $\varphi$, $\lambda$ is the Gauss-Bonnet coupling constant having  dimension of $length$ and ${\cal R}^2_{GB}$ is the Gauss-Bonnet invariant defined by ${\cal R}^2_{GB}=R^2 - 4 R_{\mu\nu} R^{\mu\nu} + R_{\mu\nu\alpha\beta}R^{\mu\nu\alpha\beta}$, where $R$ is the Ricci scalar, $R_{\mu\nu}$ is the Ricci tensor and $R_{\mu\nu\alpha\beta}$ is the Riemann tensor.  

    In order to study the problem we chose the target space to be 3-dimensional maximally symmetric space, namely $\mathbb{S}^3$, $\mathbb{H}^3$ or $\mathbb{R}^3$
    	with the metric 
    	\begin{eqnarray}
    	\gamma_{ab}(\varphi)d\varphi^a d\varphi^b= a^2\left[d\chi^2 + H^2(\chi)(d\varTheta^2 + \sin^2\varTheta d\Phi^2) \right],
    	\end{eqnarray}
    	where $a>0$ is a constant  and $\varTheta$ and $\Phi$ are the standard angular coordinates on the 2-dimensional sphere $\mathbb{S}^2$. In addition we need to specify the potential $V(\varphi)$ and the coupling function $f(\varphi)$. In this study we shall consider the simpler case with $V(\varphi)=0$.
    	
    	The three possibilities for the target space are given by the metric function $H(\chi)$: $H(\chi)=\sin\chi$ for the spherical geometry, $H(\chi)=\sinh\chi$ for the hyperbolic geometry and
    	$H(\chi)=\chi$ for the flat geometry. The parameter $a$ is related to the curvature $\kappa$ of  $\mathbb{S}^3$ and $\mathbb{H}^3$, and we have $\kappa=1/a^2$ for spherical  and $\kappa=-1/a^2$ for hyperbolic geometry. The scalar fields we chose in a nontrivial way -- only $\chi = \chi(r)$ depends on the radial coordinate $r$, while the other scalar field are given by $\varTheta = \theta$ and $\Phi = \phi$. In addition we take the coupling function $f(\varphi)$ to depend on $\chi$ only. In this way the equations for $\varTheta$ and $\Phi$ separate from the rest, and the spacetime metric will be spherically symmetric for the ansatz bellow.

	In the present paper we are interested in the static and spherically symmetric neutron star solutions to the equations of MSGB gravity
	with a metric 
	\begin{eqnarray}
	ds^2= - e^{2\Gamma}dt^2 + e^{2\Lambda}dr^2 + r^2(d\theta^2  + \sin^2\theta d\phi^2),
	\end{eqnarray} 
	where $\Gamma$ and $\Lambda$ depend on the radial coordinate $r$ only.

	 With the ansatz  for the scalar fields and using the above form of the metric, we obtain the following  reduced field equations 
	\begin{eqnarray}
	&&\frac{2}{r}\left[1 +  \frac{2}{r} (1-3e^{-2\Lambda})  \Psi_{r}  \right]  \frac{d\Lambda}{dr} + \frac{(e^{2\Lambda}-1)}{r^2} 
	- \frac{4}{r^2}(1-e^{-2\Lambda}) \frac{d\Psi_{r}}{dr} \nonumber \\ 
	&& \hspace{0.5cm} - a^2\left[ \left( \frac{d\chi}{dr}\right)^2 + 2e^{2\Lambda}\frac{H^2(\chi)}{r^2}\right] = 8\pi\rho e^{2\Lambda}, \label{DRFE1}\\ && \nonumber \\
	&&\frac{2}{r}\left[1 +  \frac{2}{r} (1-3e^{-2\Lambda})  \Psi_{r}  \right]  \frac{d\Gamma}{dr} - \frac{(e^{2\Lambda}-1)}{r^2} - a^2\left[ \left( \frac{d\chi}{dr}\right)^2 - 2e^{2\Lambda}\frac{H^2(\chi)}{r^2}\right] = 8\pi p e^{2\Lambda},\label{DRFE2}\\ && \nonumber \\
	&& \frac{d^2\Gamma}{dr^2} + \left(\frac{d\Gamma}{dr} + \frac{1}{r}\right)\left(\frac{d\Gamma}{dr} - \frac{d\Lambda}{dr}\right)  + \frac{4e^{-2\Lambda}}{r}\left[3\frac{d\Gamma}{dr}\frac{d\Lambda}{dr} - \frac{d^2\Gamma}{dr^2} - \left(\frac{d\Gamma}{dr}\right)^2 \right]\Psi_{r} 
	\nonumber \\ 
	&& \hspace{0.5cm} - \frac{4e^{-2\Lambda}}{r}\frac{d\Gamma}{dr} \frac{d\Psi_r}{dr} + a^2\left(\frac{d\chi}{dr}\right)^2 = 8\pi p e^{2\Lambda}, \label{DRFE3}\\ && \nonumber \\
	&& \frac{d^2\chi}{dr^2}  + \left(\frac{d\Gamma}{dr} \nonumber - \frac{d\Lambda}{dr} + \frac{2}{r}\right)\frac{d\chi}{dr} - \frac{2\lambda^2}{a^2r^2} \frac{df(\chi)}{d\chi}\left\{(1-e^{-2\Lambda})\left[\frac{d^2\Gamma}{dr^2} + \frac{d\Gamma}{dr} \left(\frac{d\Gamma}{dr} - \frac{d\Lambda}{dr}\right)\right]   \right. \nonumber \\
	&& \left. \hspace{0.5cm}  + 2e^{-2\Lambda}\frac{d\Gamma}{dr} \frac{d\Lambda}{dr}\right\} =  \frac{2}{r^2} H(\chi)\frac{dH(\chi)}{d\chi}e^{2\Lambda} \label{DRFE4}
	\end{eqnarray} 
	with 
	\begin{eqnarray}
	\Psi_{r}=\lambda^2 \frac{df(\chi)}{d\chi} \frac{d\chi}{dr}.
	\end{eqnarray}
	
The above dimensionally reduced field equations we have to supplement with the equation for the hydrostatic equilibrium of the fluid

\begin{eqnarray} \label{Eq:dp}
\frac{dp}{dr}= - (\rho + p) \frac{d\Gamma}{dr}.
\end{eqnarray}	

\section{Numerical setup and results}	

In the present paper we will present results for one realistic equation of state (EoS), namely MPA1 \cite{Muther1987}. This EoS allows maximal masses well above two solar masses, and it is in agreement with the constraints set by the observations of double neutron star mergers \cite{Abbott2017a}. For the numerical implementation we are adopting its piecewise polytropic approximation \cite{Read2009}.

We have studied different  coupling functions $f(\chi)$ which allow spontaneous scalarization, namely functions which fulfill the condition $\frac{df}{d\chi}(0) = 0$. For some of those functions we found only unstable solutions (energetically less favorable than the GR ones), and for some of them we were unable to find solutions at all. In the present study we are using the following one 

\begin{equation}
f(\chi) = -\frac{1}{2\beta}\left(1-e^{-\beta\chi^2}\right),
\end{equation}
where $\beta$ is a positive parameter. The coupling constant $\lambda$,  is presented in dimenstionless units 

\begin{equation}
\lambda \rightarrow \frac{\lambda}{R_0}
\end{equation}
where $R_0 = 1.476$ km is one half of the gravitational radius of the Sun.

Finding the scalarized solutions and computing the scalarized branches is difficult and time consuming task. This is why we constrained ourselves to only one equation of state and one coupling function with our main task being to construct representative solutions and determine their general properties and dependence from the parameters in the theory. 

The reduced field equations (\ref{DRFE1})-(\ref{DRFE4}) and the equations for the hydrostatic equilibrium (\ref{Eq:dp}) are solved numerically with the natural boundary conditions -- regularity at the center of the star and asymptotic flatness at infinity:

\begin{equation}
\Lambda(0) = 0, \quad \frac{d\Gamma}{dr}(0) = 0, \quad \chi(0) = 0,
\end{equation}
and

\begin{equation}
\Lambda|_{r\rightarrow\infty} \rightarrow 0, \quad \Gamma|_{r\rightarrow\infty} \rightarrow 0, \quad \chi|_{r\rightarrow\infty} \rightarrow 0.
\end{equation}

The asymptotic behavior of the scalar field can be derived from the field equations to be $\chi \sim 1/r^2$, therefore the scalar field has no scalar charge. This has strong physical consequences for the theory. The absence of scalar charge means the dipole scalar radiation in binary system will be strongly suppressed, therefore no strong constraints can be set on the theory parameters by the observations of binary pulsars.  

For the coupling function we exploit, the trivial scalar field $\chi = 0$ is always a solution of the field equations (\ref{DRFE1})-(\ref{DRFE4}), and the results coincide with the general relativity ones. From now on trivial solutions and GR solutions will be used interchangeably. In this paper we are searching the three dimensional parameter space $(a^2,\lambda,\beta)$ for bifurcation points from which new branches of solutions with nontrivial scalar field emerge from the trivial branch. This task turned out to be nontrivial. On one hand is the number of free parameters in the theory, and on the other -- the severe numerical difficulties emerging from the system (\ref{DRFE1})-(\ref{DRFE4}) by itself.  

Concerning the different target spaces, we found solutions for all three of them. The differences between them, however, are negligible, and can not be seen when comparing the results on the figures.  The absence of deviations we explain with the small values of the scalar field. We found that even for the maximal observed values of the scalar fields, the values of the metric functions $H(\chi) = \sin(\chi)$, $H(\chi) = \chi$ and $H(\chi) = \sinh(\chi)$ are practically indistinguishable. This is why we chose to present only the results for the spherical target space with $H(\chi) = \sin(\chi)$.

\begin{figure}[]
	\centering
	\includegraphics[width=0.45\textwidth]{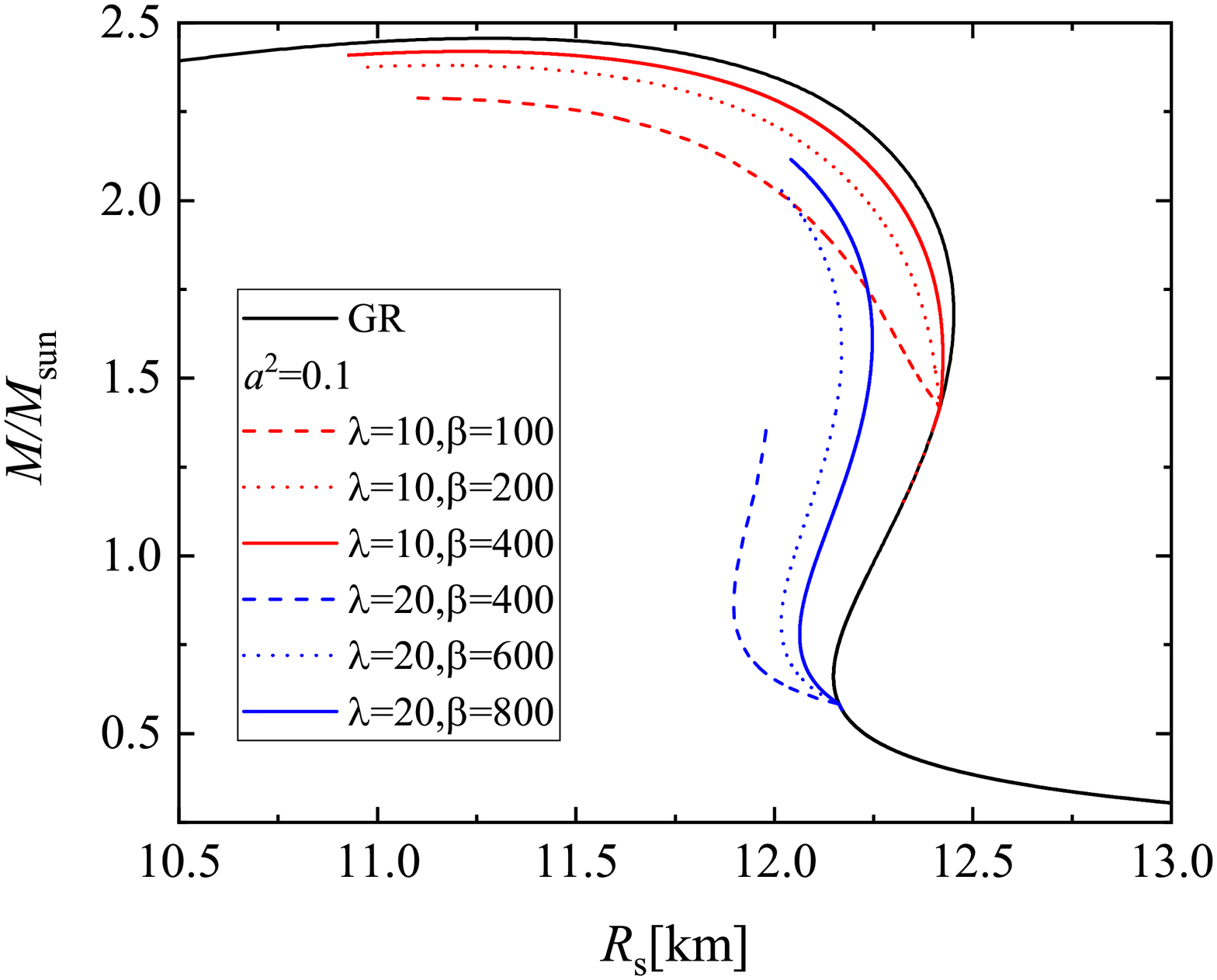}
	\includegraphics[width=0.45\textwidth]{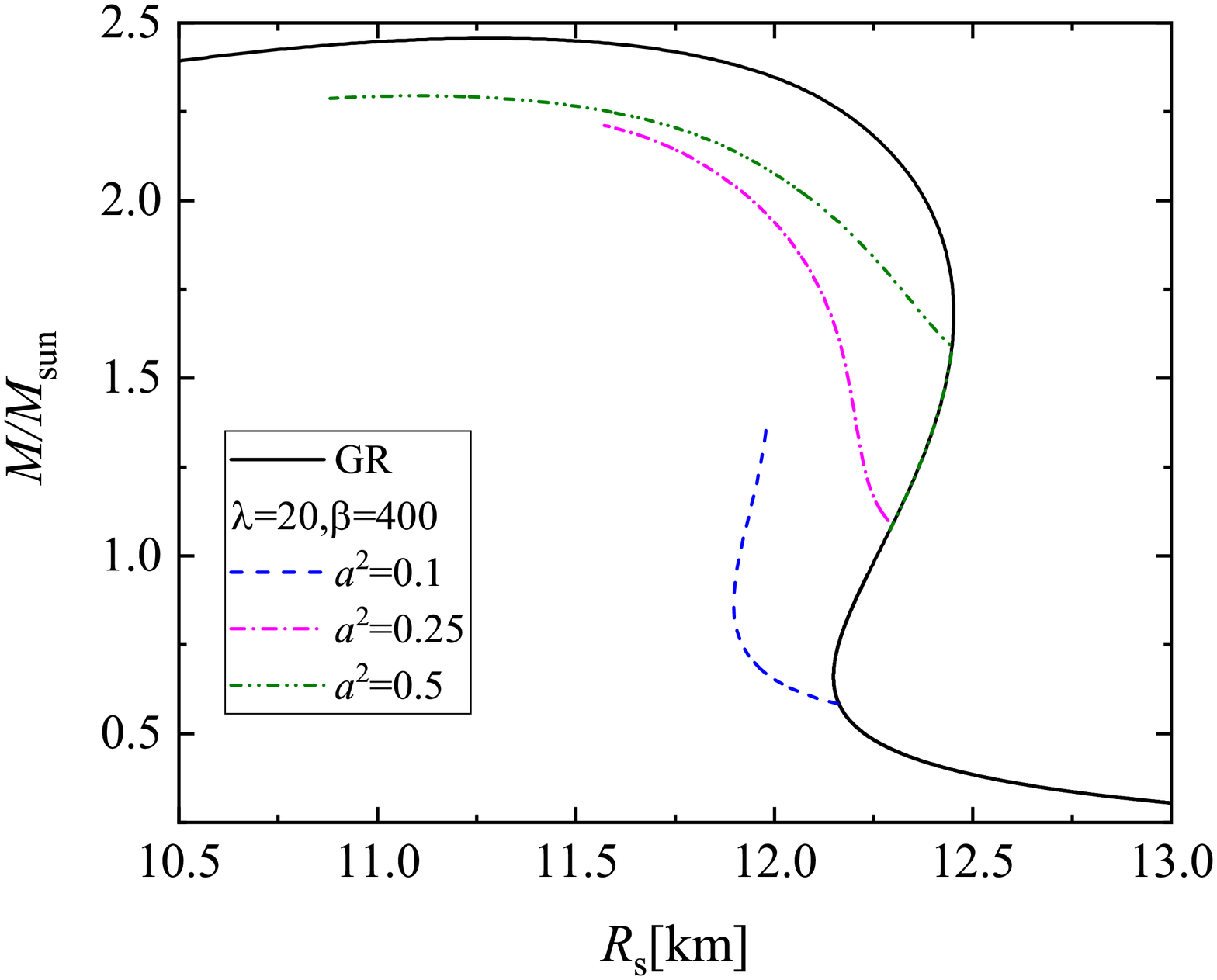}
	\caption{Mass of radius relation for models with different values of the parameters. \textit{Left} Fixed value of $a^2$ and different values of $\lambda$ and $\beta$.  \textit{Right} Fixed values of $\lambda$ and $\beta$ and different values of $a^2$. The mass of the star is in solar masses and the radius -- in kilometers. }
	\label{Fig:M_R}
    \end{figure}

In Fig. \ref{Fig:M_R} we present the mass of the star, in solar masses, as a function of its radius, in km. In the left panel the value of $a^2$ is fixed and different combinations of $\lambda$ and $\beta$ are studied. In the right panel $\lambda$ and $\beta$ are fixed and different values for $a^2$ are studied. On both panels it is clear that the results are qualitatively similar to the results in pure Gauss-Bonnet gravity presented in \cite{Doneva2018a}.
The nontrivial solutions emerge from the bifurcation points and the branches are terminated either after they reach maximal mass or at some smaller mass due to numerical difficulties. For all sets of parameters the computation procedure gets more sensitive to the initial conditions with the increase of the central energy density, and the computational time for the individual models continuously increases along the branch. Hence, some of the branches were terminated before the maximal mass, at some reasonable computational time. The maximal mass, when it is reached, is always lower than the GR one.  However, we found that for the same set of parameters, $\lambda$ and $\beta$, a value for $a^2$ could be chosen (smaller than unity) so that the GB and the MSGB solutions bifurcate at the same central energy density. In this case the MSGB neutron stars have smaller masses compared to the GB ones. 

On the left panel of the figure one can see that for a fixed value of $a^2$ the bifurcation point moves to higher masses (equivalently -- central densities) with the decrease of $\lambda$. For fixed $\lambda$ the nontrivial solution branch gets longer and tends to the GR one with the increase of $\beta$ (the bifurcation point does not move). In the limiting case $\beta \rightarrow \infty$ the nontrivial branch will coincide with the trivial one. On the right panel one can see that for fixed $\lambda$ and $\beta$ the bifurcation point moves to higher masses, and the branch gets longer with the increase of $a^2$. We were not able to reach $a^2 = 1$, though.

The behavior described above may pose some constrains on the free parameters in the theory.  If the equation of state allows maximal masses higher than two solar masses in GR, the parameters of the MSGB theory should be chosen in such a way that the maximal mass is above two solar masses. This, however, is not very restrictive, because two of the parameters allow significant deviations in the bifurcation point and the case $\beta \rightarrow \infty$ always tends to the trivial solution no matter what are the values of the other two parameters. In addition, not all branches reach maximal masses.   

For all presented combination of parameters we were able to find branches of solutions with different number of zeroes of the scalar field -- no zeroes, one zero, etc. We found that the bifurcation point moves to higher masses and the branch gets shorter with the increase of the number of zeroes of the field. The presented in this paper results are only for solutions with no zeroes. Those solutions, as explained bellow, are the stable ones. 

\begin{figure}[]
	\centering
	\includegraphics[width=0.45\textwidth]{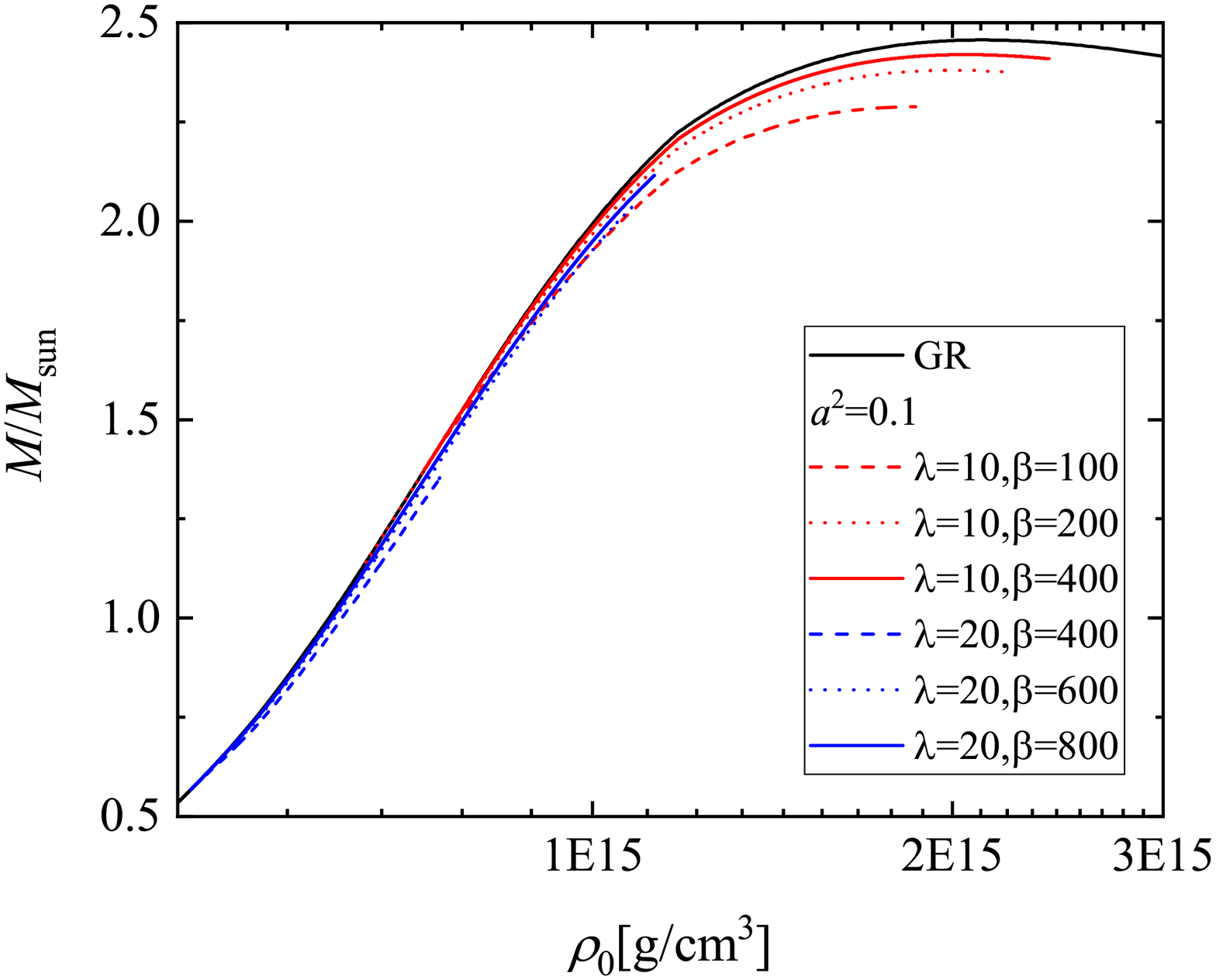}
	\includegraphics[width=0.45\textwidth]{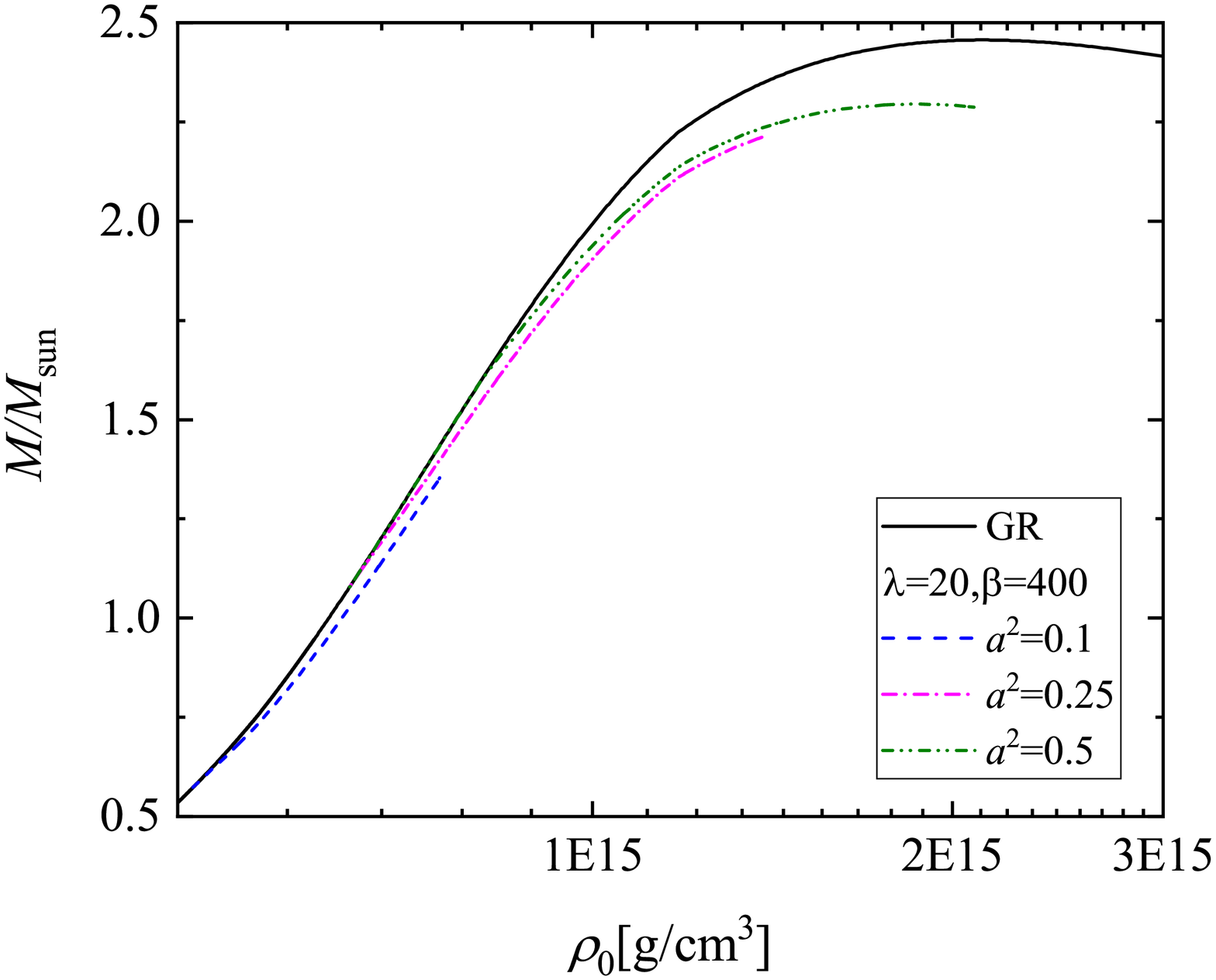}
	\caption{Mass of the star as a function of the central energy density for models with different values of the parameters. \textit{Left} Fixed value of $a^2$ and different values of $\lambda$ and $\beta$.  \textit{Right} Fixed values of $\lambda$ and $\beta$ and different values of $a^2$. The mass of the star is in solar masses and the energy density -- in $g/cm^3$. }
	\label{Fig:M_rho}
\end{figure}

In Fig. \ref{Fig:M_rho} we plot the mass of the neutron stars as a function of the central energy density. The presented models in the left and in the right panel correspond to those in Fig. \ref{Fig:M_R}. For all models the deviations from GR are relatively small.

\begin{figure}[]
	\centering
	\includegraphics[width=0.45\textwidth]{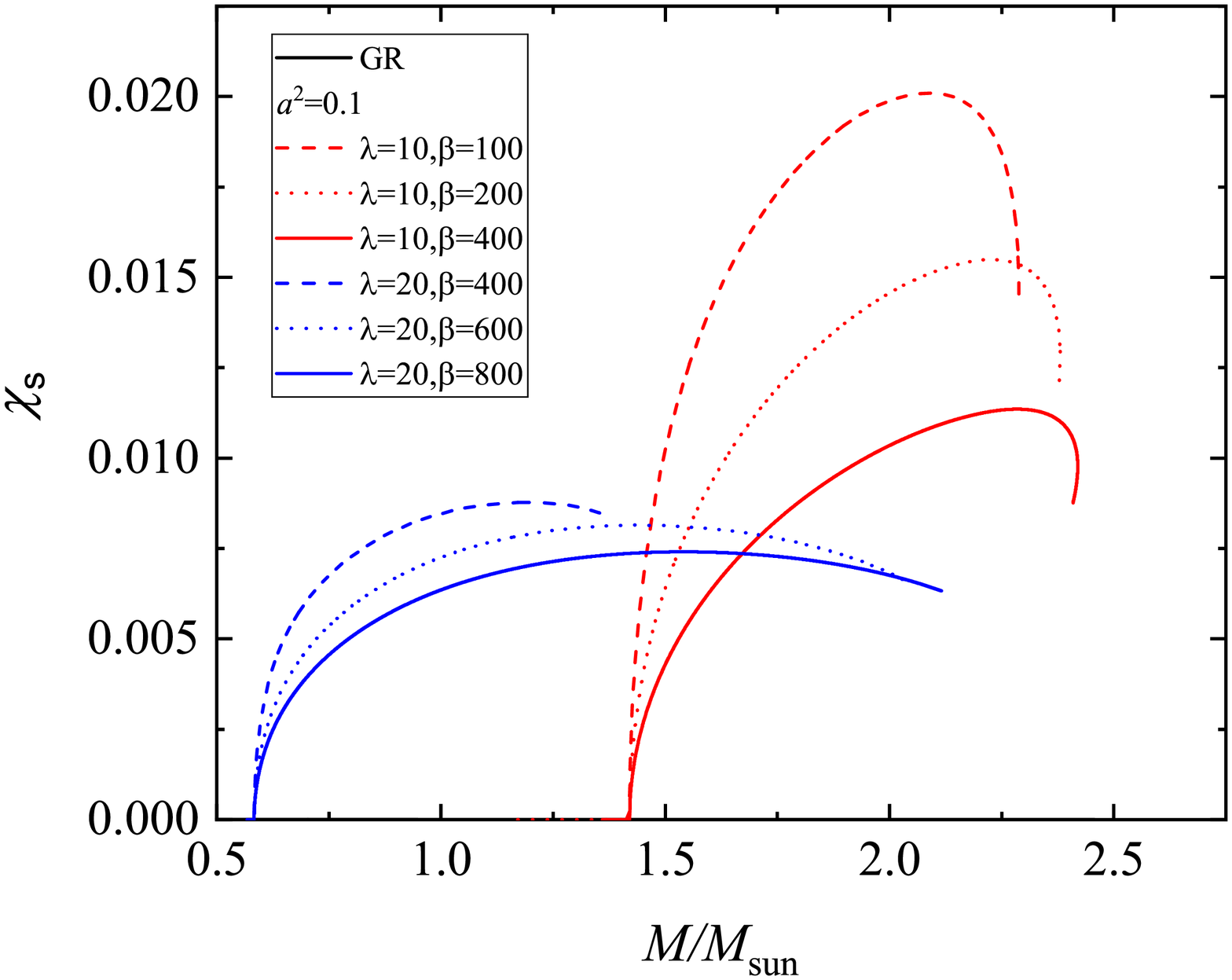}
	\includegraphics[width=0.45\textwidth]{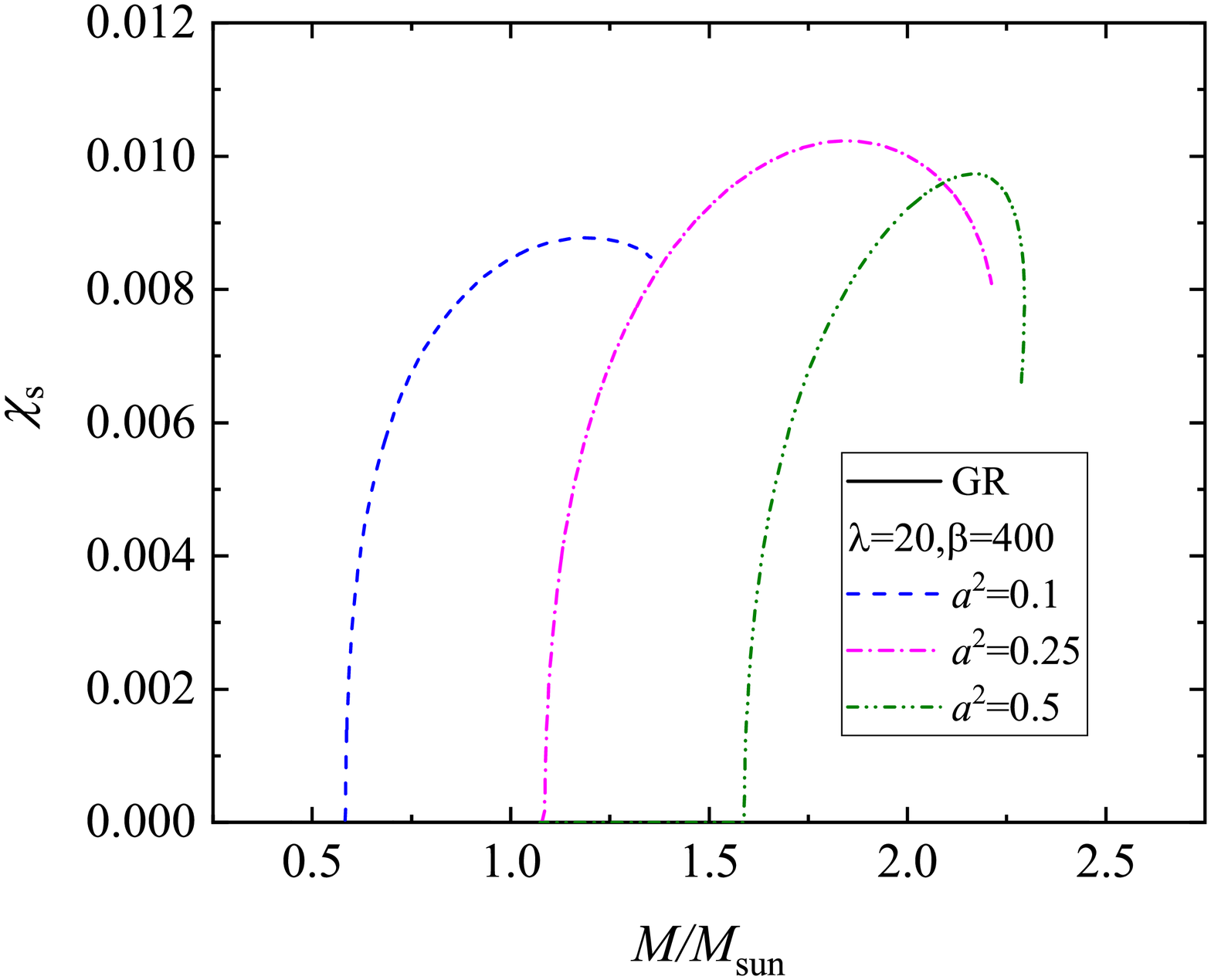}
	\caption{The value of the scalar field on the surface of the star as a function of the mass of the star. \textit{Left} Fixed value of $a^2$ and different values of $\lambda$ and $\beta$.  \textit{Right} Fixed values of $\lambda$ and $\beta$ and different values of $a^2$. The mass of the star is in solar masses. }
	\label{Fig:chiS_M}
    \end{figure}

Due to the absence of scalar charge, in Fig. \ref{Fig:chiS_M} we plotted the value of the scalar field on the surface $\chi_S$ of the star as function of the NS mass. $\chi_S$ is always higher for smaller values of $\lambda$ and decreases with the increase of $\beta$. In the same time $a^2$ does not have any significant effect on the maximal value of $\chi_S$.

\begin{figure}[]
	\centering
	\includegraphics[width=0.45\textwidth]{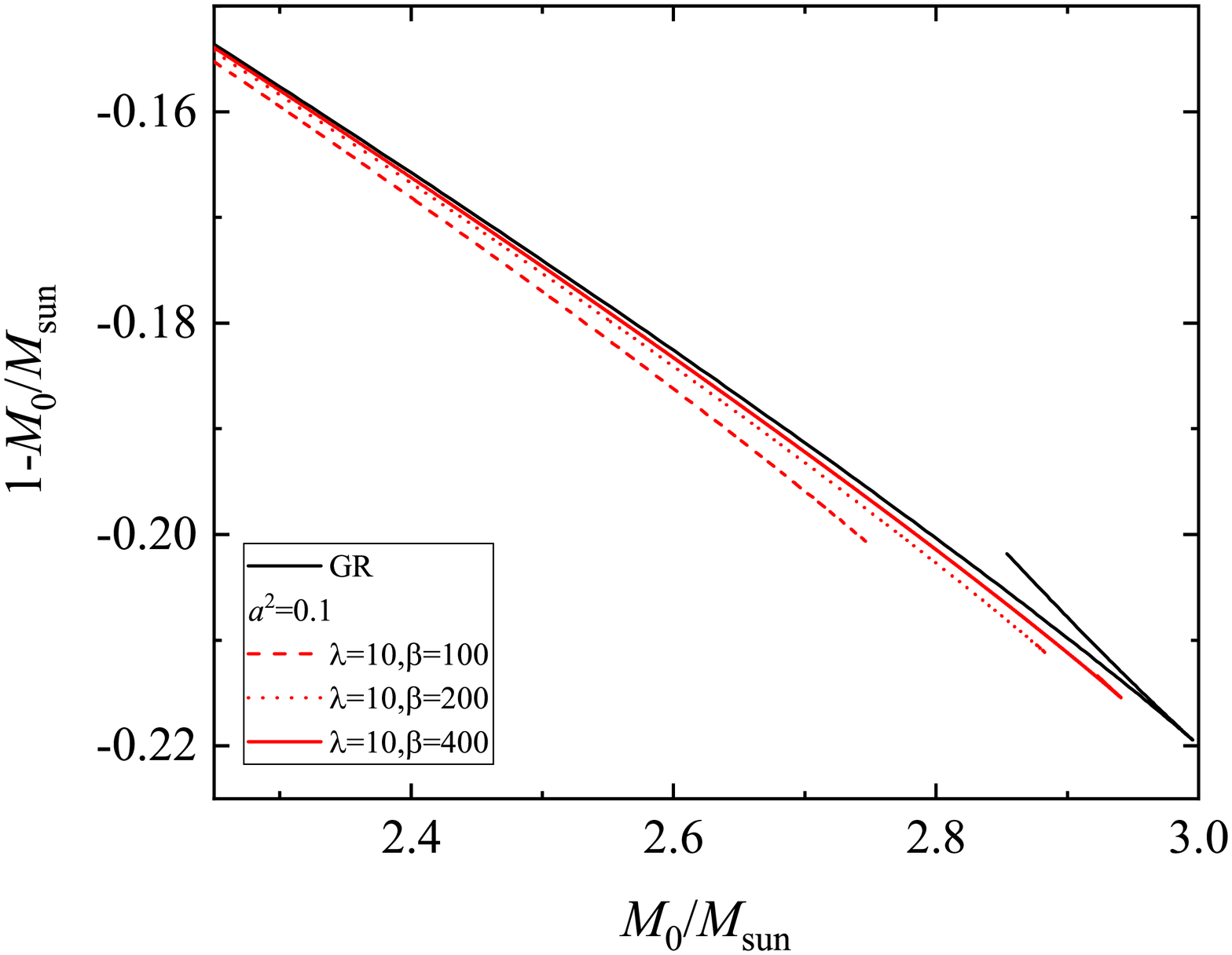}
	\includegraphics[width=0.45\textwidth]{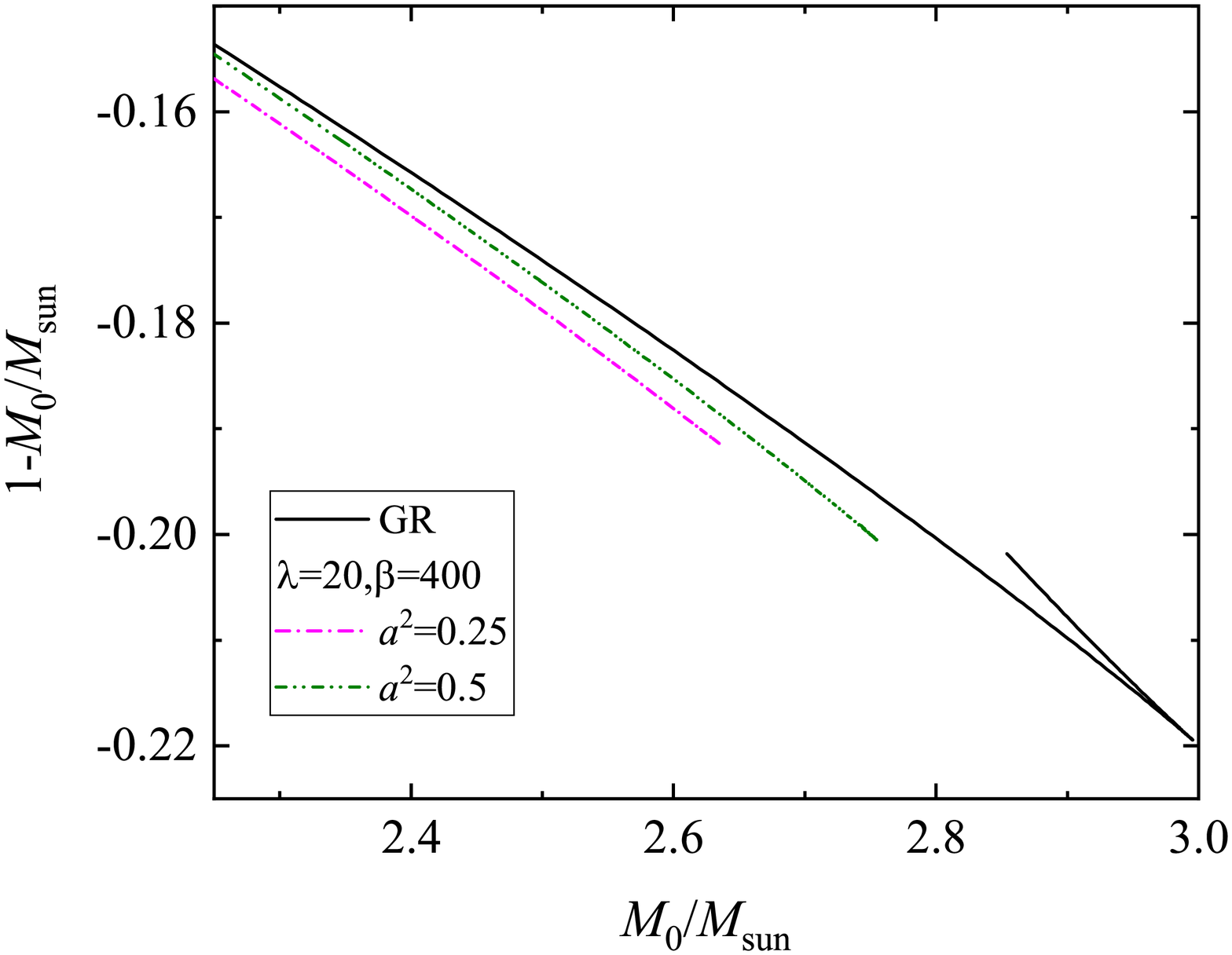}
	\caption{The binding energy $1-M_0/M_{\rm{sun}}$ as a function of the rest mass of the star $M_0$. \textit{Left} Fixed value of $a^2$ and different values of $\lambda$ and $\beta$.  \textit{Right} Fixed values of $\lambda$ and $\beta$ and different values of $a^2$. The rest mass of the star is in solar masses.  }
	\label{Fig:BE}
    \end{figure}

As an indication for the stability of the neutron star solutions, in Fig. \ref{Fig:BE} we present the binding energy of the star, $1-\frac{M_0}{M_{\rm{sun}}}$, as a function of $M_0/M_{\rm{sun}}$, where $M_0$ is the rest mass of the star. For better visualization, only some of the parameters studied above are presented. The MSGB solutions have higher (by absolute value) binding energy compared to the GR ones, which makes them energetically more favorable. The cusp in the GR solution marks the unstable models in the mass-energy density relation. In the MSGB case, such cusp is present as well for the sets of parameters for which the maximal mass can be reached. The models with zeros of the scalar field have lower (by absolute value) binding energy compared to the corresponding ones with no zeros. Therefore, by analogy with other STTs, the solutions with zeros of the scalar field are most probably unstable. However, the binding energy is only an indicative sign for stability, and the radial perturbations of those models should be studied for definitive answer.    

\section{Conclusion}

    In the present paper we have constructed scalarized neutron stars in multi-scalar Gauss-Bonnet theory with maximally symmetric target space and nontrivial map $\varphi: spacetime \rightarrow targer\ space$, compatible with spherical symmetry and explicitly given by $\varphi = (\chi(r), \varTheta = \theta, \Phi = \phi)$. We have studied the simplest three possible choices for maximally symmetric target space, namely $\mathbb{S}^3$,  $\mathbb{H}^3$, and  $\mathbb{R}^3$. The models we have examined have vanishing scalar field in the center of the star, and the presented results are for coupling function which allows scalarization. Regarding the target space, no differences between the solutions in the three cases were observed.   This is easily explained by the low values of the scalar field, which makes the metric function $H(\chi)$, which specifies the target space, effectively indistinguishable between the three cases. Hence, the presented results were only for the spherical target space.    
    
    The obtained neutron star solutions look qualitatively similar to the neutron star solutions in the pure GB gravity. The branches with nontrivial scalar field bifurcate from the trivial solution, and the maximal mass of the scalarized solutions is always lower than the GR one.  For the same sets of values for the parameters in the theory, multiple branches of solutions exist, characterized by the number of zeros the scalar field has. The bifurcation point, though, depends strongly on the value of the coupling constant $\lambda$, the parameter $a^2$ in the target space metric and the number of zeroes of the scalar field. 
    The maximal mass of the MSGB solutions is always lower compared to the GR one which may allow some constraints to be imposed on the parameters of the theory. In the same time, however, due to numerical difficulties not all branches of solutions reach maximal mass. 
    
    In order to obtain some indications about the stability of the scalarized models we studied the binding energy of the star $1 - M_0/M_{sun}$. The scalarized solutions have higher (by absolute value) binding energy than the GR ones which makes them energetically more favorable. Concerning the branches with different numbers of zeros of the scalar field, the branch with no zeros has highest (by absolute value) binding energy, therefore most probably this branch is stable and the rest of the branches -- unstable. This conclusion is based on the binding energy and the behavior of the neutron star solutions in similar theories. For definitive answer about the stability of the models, however, the radial perturbations should be studied.   
	
	\section*{Acknowledgements}
    KS is supported 	by the Bulgarian national program "Young Scientists and Postdoctoral Research Fellows 2021". Networking support by the COST action CA16214 is gratefully  acknowledged.
	
	
	\bibliography{references}

\end{document}